\documentclass[12pt,onecolumn,journal]{IEEEtran}
\usepackage{amsmath,amsfonts}
\usepackage{algorithmic}
\usepackage{algorithm}
\usepackage{array}
\usepackage[caption=false,font=normalsize,labelfont=sf,textfont=sf]{subfig}
\usepackage{textcomp}
\usepackage{stfloats}
\usepackage{url}
\usepackage{verbatim}
\usepackage{graphicx}
\usepackage{cite}
\begin{document}

\title{Pricing for Reconfigurable Intelligent Surface Aided Wireless Networks: Models and Principles}

\author{Yulan Gao, \IEEEmembership{}
Yue Xiao, \IEEEmembership{Member,~IEEE,}
Xianfu Lei, \IEEEmembership{Member,~IEEE,}
Qiaonan Zhu, \IEEEmembership{}
Dusit Niyato, \IEEEmembership{Fellow,~IEEE,}
Kai-Kit Wong, \IEEEmembership{Fellow,~IEEE,}
Pingzhi Fan, \IEEEmembership{Fellow,~IEEE,}
and
Rose Qingyang Hu \IEEEmembership{Fellow,~IEEE}

\thanks{Yulan Gao is with the National Key Laboratory of Science and Technology on Communications, University of Electronic Science and Technology of China, Chengdu 611731, China, and also with the School of Computer Science and Engineering, Nanyang Technological University Singapore 639798.

Yue Xiao and Qiaonan Zhu are with  the National Key Laboratory of Science and Technology on Communications, University of Electronic Science and Technology of China, Chengdu 611731, China. The corresponding author is Yue Xiao (email:  xiaoyue@uestc.edu.cn).

Xianfu Lei is with the School of Information Science and Technology, Southwest Jiaotong University, Chengdu 610031, China, and also with the National Mobile Communications Research Laboratory, Southeast University, Nanjing 210096, China (e-mail: xflei@swjtu.edu.cn).

Dusit Niyato is the School of Computer Science and Engineering, Nanyang Technological University Singapore 639798.

Kai-Kit Wong iswith the Department of Electronic and Electrical Engineering, University College London, London WC1E 6BT, U.K.

Pingzhi Fan is with the Provincial Key Laboratory of Information Coding and Transmission, Southwest Jiaotong University, Chengdu 611756, China.

Rose Qingyang Hu is with the Department of Electrical and Computer Engineering, Utah State University, Logan, UT 84322 USA.
 }
}

\markboth{~}%
{Shell \MakeLowercase{\textit{et al.}}: A Sample Article Using IEEEtran.cls for IEEE Journals}


\maketitle

\begin{abstract}
Owing to the recent advancements of meta-materials and meta-surfaces, the concept of reconfigurable intelligent surface (RIS) has been embraced to meet the spectral- and energy-efficient, and yet cost-effective solutions for the sixth-generation (6G)  wireless networks.
From an operational standpoint, RISs can be easily deployed on the facades of buildings and indoor walls.
Albeit promising, in the actual network operation, the deployment of RISs may face challenges because of the willingness and benefits of RIS holders from the aspect of installing RISs on their properties.
Accordingly, RIS-aided wireless networks are faced with a formidable mission: how to balance the wireless service providers (WSPs) and RIS holders in terms of their respective interests.
To alleviate this deadlock, we focus on the application of pricing models in RIS-aided wireless networks in pursuit of a win-win solution for both sides.
Specifically, we commence with a comprehensive introduction of RIS pricing  with its potential applications in RIS networks, meanwhile the fundamentals of pricing models are summarized in order to benefit both RIS holders and WSPs.
In addition, a Stackelberg game-based model is exemplified to illustrate the operation of utility-maximization pricing.
Finally, we highlight open issues and future research directions of applying pricing models to the RIS-aided wireless networks.

\end{abstract}
\begin{IEEEkeywords}
Reconfigurable intelligent surface, resource management, pricing models, 6G.
\end{IEEEkeywords}

\section{Introduction}
\IEEEPARstart{W}{ith} the commercialization of the fifth generation (5G) wireless network and the exploration and development of its applications in vertical industries, the vision of the sixth generation (6G)  wireless network has gradually  attracted wide attentions.
6G wireless networks introduce new application scenarios while
proposing higher performance indicators, such as seamless global connectivity, higher spectral- and energy- efficient, ultra-reliable communications, and security, etc.
In conventional networks, the transceiver module is a symmetrical architecture with independent radio frequency (RF) chains with high energy consumption components such as power amplifiers.
Nevertheless, a large number of accessed devices in 6G networks will inevitably lead to a sharp increase in power consumption.
Therefore, the realization of high data rate with significantly reduced energy consumption and implementation cost for 6G wireless networks is still imperative.
\begin{figure}[!t]
\centering
\includegraphics[width=1\linewidth]{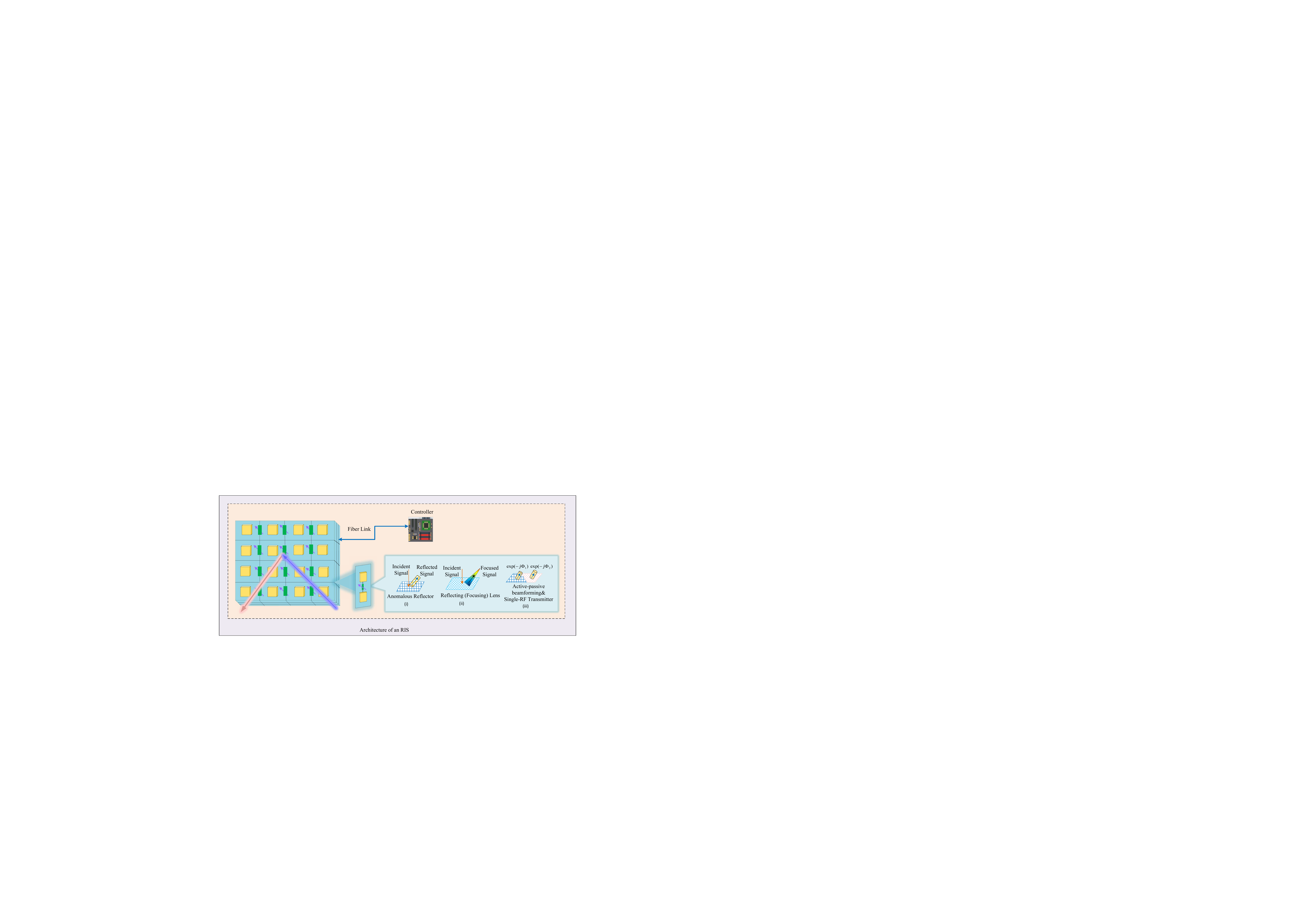}
\caption{Prototype/schematic of the RIS-aided wireless communication.}
\label{fig:1}
\end{figure}
\begin{figure*}[htbp]
\centering
\includegraphics[scale=.28]{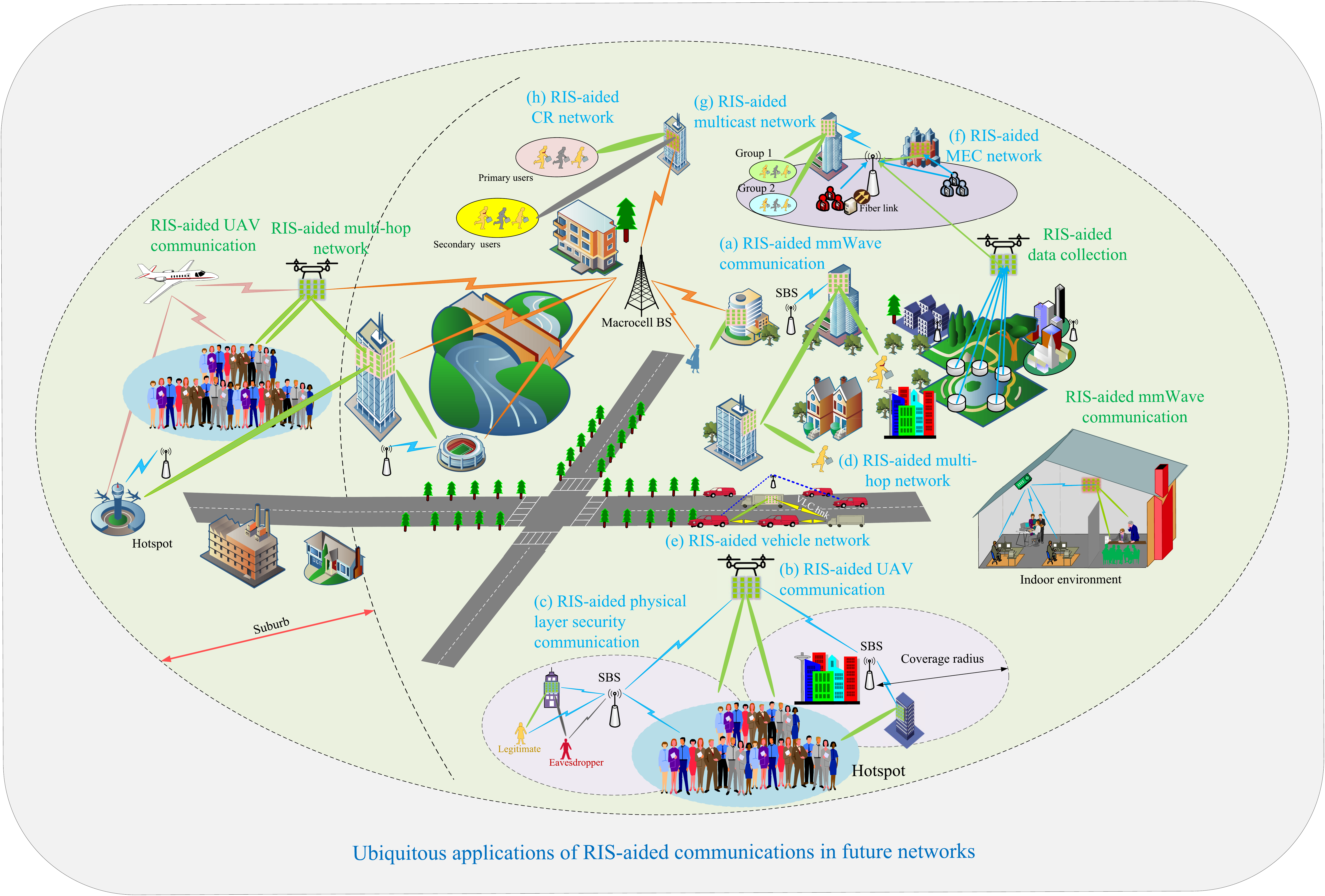}
\caption{Ubiquitous applications of RIS-aided communication systems in future wireless networks.}
\label{fig:2}
\end{figure*}

Recently, reconfigurable intelligent surface (RIS) has been  promoted as a promising technology to break the above-mentioned deadlock, driven by the artificial intelligence (AI) theory, emerging meta-materials, and integrated antenna technologies \cite{liu2021reconfigurable}.
Benefiting from the breakthrough on the programmable meta-material, RIS was speculated as one of the key enabling technologies for the future 6G wireless networks to construct a smart radio environment.
The prototype of RIS-aided wireless communication is shown in Fig.\ref{fig:1}, and there are four types of use according to their individual functions and application scenarios \cite{gao2020reconfigurable,di2020reconfigurable}.
In particular, {\em anomalous reflection/transmission} as shown in Fig. \ref{fig:1}(i) is configurable for specific direction, regardless of the channel attenuation and user locations.
As described in Fig. \ref{fig:1}(ii), {\em beamforming/confusing} usage type is designed for certain area with the challenge for optimizing the transmission according to the channel fading and the locations of receivers.
The last two types shown in Fig. \ref{fig:1}(iii) are  {\em joint active-passive beamforming} and {\em single RF multi-stream transmitter design}, respectively.
The former focuses on  the highest system performance via jointly active-passive beamforming design, while only one single RF chain is desired in the latter with low implementation cost by considering RIS as a special modulator.
In summary, RIS can proactively modify the wireless environment via adjusting its capacitance, resistance, and inductance.
Consequently, the energy efficiency, throughput, and coverage of networks can be significantly enhanced by manipulating the wireless signals.

In the vision of 6G, RISs will be deployed ubiquitously due to the request for  seamless coverage.
Consequently,  in actual operation, the diversity of 6G entities with different objectives may introduce challenges for  RISs deployment and multi-resource management.
The reasons are described as (i) the diverse and dense deployment of wireless devices and RISs;
(ii) the willingness of holders to install large RIS;
(iii) the inherent constraints of the service range of RIS;
(iv) a large number of users/stakeholders and RIS holders with different objectives.
In general, the inherently heterogeneous nature of an ecosystem in which traditional wireless networks and RISs coexist is intuitively attributed to the different dimension of components.
More precisely, in RIS-aided wireless networks, the fundamental infrastructures such as base stations (BSs) and RISs may belong to or be operated by different entities, such as WSPs and RIS holders, resulting in them usually having different objectives and constrains.
Accordingly, system optimization for a single objective may fail to model and determine an optimal interaction among these self-interested and rational entities.
On the other hand, recently, pricing mechanisms have been widely
developed and adopted as useful tools to address resource management issues in heterogeneous networks, whose  components are with different objectives, such as high data rate, low latency, and revenue maximization.
In the pricing process, each entity reaches the best decision according to the equilibrium analysis via the interactions among all the entities.
Accordingly, the inherent and partial information exchange nature of pricing  makes it suitable for the autonomous decision made in a distributed method.
In short, the above mentioned mechanism may be applicable for the RIS-aided wireless network which consists of a volume of autonomous entities, meanwhile meeting the demand for diverse objectives of a large number of entities in RIS-aided wireless networks.
As such, the purpose of this article is to draw attention to and spur activities on this new research direction.

In this article we focus particularly on the applications of pricing models in RIS-aided wireless networks.
Specifically, we first highlight the factors that make pricing imperative for RIS-aided wireless networks, and then review major approaches of pricing.
Specifically, two major directions are presented, i.e., the heterogeneous architecture of RIS-aided wireless networks and proper pricing of RIS.
Next, we present a demonstrative pricing model based on Stackelberg game theory to study RIS service competition.
Finally, open research directions are outlined.

\section{RIS-Aided Wireless Networks: Pricing  and Major Approaches}

\subsection{What and Why Pricing Approaches}

{\bf Heterogeneity of RISs in Future Wireless Networks:}
As shown in Fig. \ref{fig:2}, RISs will eventually pervade almost all application scenarios in the future-generation wireless networks.
For instance, RIS-aided mmWave networks are illustrated in case (a), where RIS is capable of enhancing the QoS and coverage in mmWave communication environments as diversely as indoor, open rural, suburban, and urban areas.
This is achieved by overcoming or at least mitigating the problems imposed by the severe path loss associated with the above-mentioned diverse communication environments.
Case (b) shows three dimensional networking architectures enabled by aerial RIS, where RIS is mounted on unmanned aerial vehicles (UAVs), so as to enable intelligent reflection to come from the sky.
Case (c) exhibits the use of RIS for improving the physical layer security (PLS), where RIS is capable of mitigating the information leakage.
This is achieved by deploying RIS in the vicinity of the eavesdropper, thus the reflected signal by RIS can be weakened at the eavesdropper.
In a nutshell, the ubiquitous applications of RISs illustrate  their indispensability in future wireless communications.
Obviously, the heterogeneity of RIS-aided wireless networks are determined by the inherently heterogeneous nature of 6G.
This means that the adoption of the emerging RIS technology introduces challenges for multiple resource management such as beamforming, site selection, spectrum allocation, and user schedule.
These challenges result from two dimensions, that is, 6G infrastructures (e.g., dense deployment of wireless devices, the coverage and data rate nonuniform of base stations, and the limitations of capacities) \cite{luong2018applications} and the incorporation of RISs into wireless networks (e.g., appropriate site selection, willingness of RIS holders,  the service constraints of RIS, and numerous stakeholdes with diverse objectives).
The traditional resource management solutions focus exclusively on entire system maximization, which relies on a centralized entity, resulting in  considerable information exchange between users and the network operators.
However, in RIS-aided wireless networks, not only the dimension of resource is increased, but also entities have multiple roles.
Therefore, the traditional resource management strategies are not suitable for the complex RIS-aided wireless networks.

\begin{figure}[!t]
\centering
\includegraphics[width=1\linewidth]{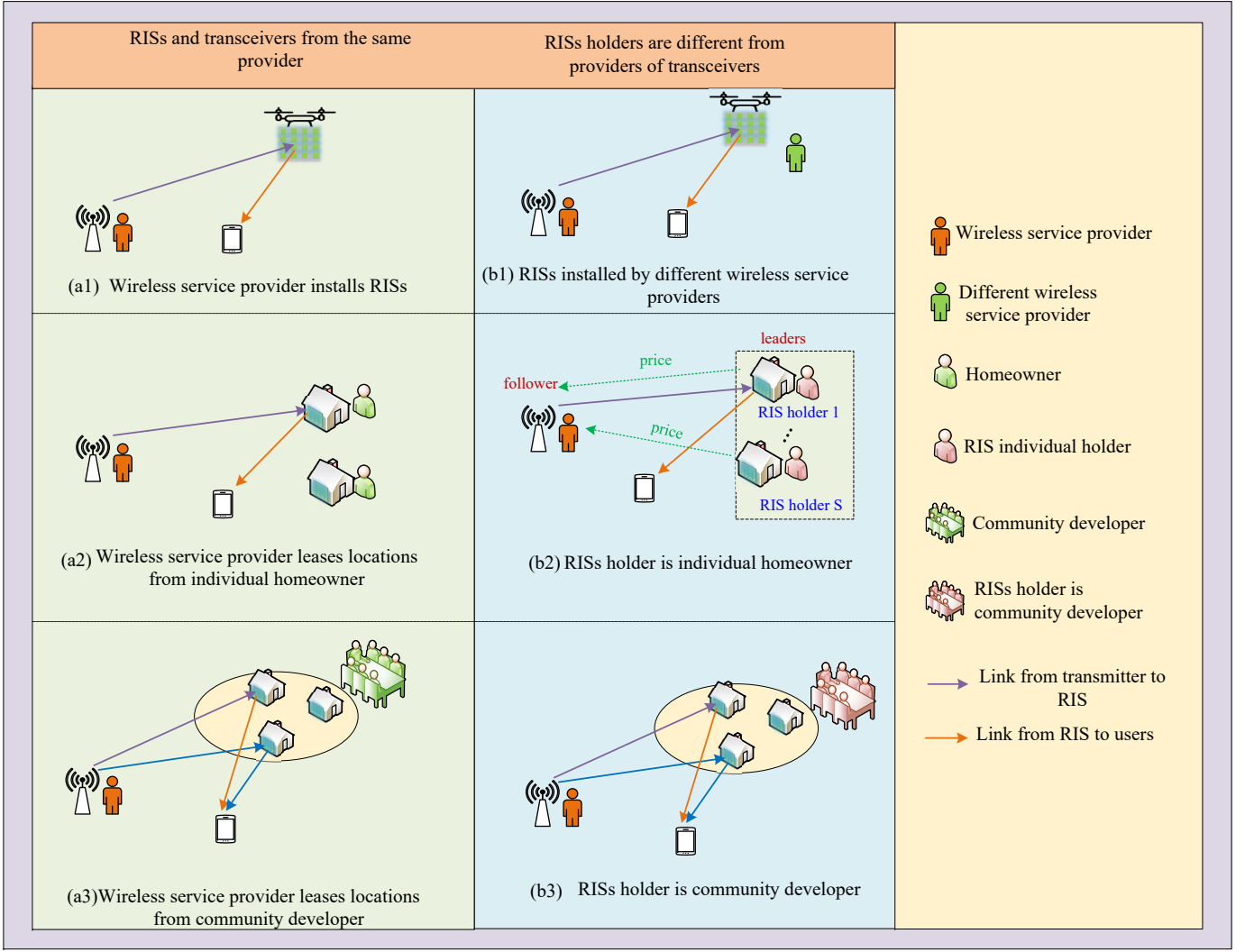}
\caption{The key different relationships between RIS holder and wireless service provider.}
\label{fig:3}
\end{figure}

{\bf Multiple Entities and Rationality:}
The heterogeneity of an ecosystem in which traditional wireless networks and RISs coexist is intuitively attributed to that they may belong to or being operated by different entities, such as RIS holders, WSPs, and community developers respectively with their individual interests and limitations.
The key different relationships between the RIS holder and WSP are shown in Fig. \ref{fig:3}, which includes two major categories:
(a) RISs and infrastructure/transceivers from the same WSP;
(b) RIS holders different from service providers.
In particular, there are three circumstances in the former type:
(a1) RISs are installed on the devices deployed by the WSP;
(a2) RISs are installed on walls or other places that WSP rents from homeowners;
(a3) wireless service providers lease positions from developers to deploy RISs for services.
Correspondingly, the latter is further subdivided into three cases:
(b1) RISs are installed on the devices deployed by another WSP;
(b2) RISs are installed by individual homeowners;
(b3) RISs holders are community developers.
The traditional methods, e.g., the system optimization via active-passive beamforming design, can provide optimal resource allocation for the RIS-aided wireless network.
However, they usually support a unilateral objective and thus may fail to model and determine an optimal interaction among these self-interested and rational entities.
Therefore, the traditional methods may not be suitable for the RIS-aided wireless network, especially when the rationality of RIS holders and WSPs is considered as the most important factor.
Meanwhile, pricing approaches have been recently developed and adopted as useful tools to address resource management in the 5G wireless networks, where different wireless components also have different objectives, such as high data rate, low latency, and revenue maximization \cite{niyato2016economics}.
\subsection{Major Approaches of Pricing}

In this subsection, two major approaches for pricing based on optimization formulation and game theory are discussed.

{\bf Optimization-based Pricing:}
As widely accepted, an optimization formulation can be used to obtain the price strategy for a given objective and a set of constraints, when there is only one WSP.
The authors of \cite{kelly1998rate} have proffered price-based rate allocation to maximize the system utility, as a groundbreaking work.
An established theoretical tool for this kind of problems is provided by the theory of dual-primal method.
More precisely, the optimization formulation has been decomposed into multiple sub-problems for optimizing the network and user utility respectively, which is built on the premise of constructing rate as a function of price.
In line with this work,  a vast corpus of literature has focused on developing optimization-based pricing strategies for rate adaptation, access control, and radio resource management, etc.
For instance, Ref. \cite{lee2005downlink} has investigated pricing-based power allocation for code division multiple access (CDMA) systems, in which the utility of user is considered as a mapping of the signal-to-noise ratio (SINR) and the price charged for per unit of the transmit power.
Ref. \cite{badia2006pricing} has studied a radio resource management and pricing scheme for multimedia wireless local area networks.
Again, solutions that focus exclusively on the objective of the entire system are not aligned with the individual interests of the entities, which, as we mentioned above, are crucial for the deployment and operations in RIS-aided wireless networks.

{\bf Game-Theory-Based Pricing:}
Game theory has propelled to the forefront of investigating  the interaction among multiple entities in a wireless network \cite{niyato2007wireless}.
Compared to optimization-based approaches, game-theoretic formulations focus on providing individual optimal solutions rather than system optimal utility.
According to the corresponding incentive mechanism design, there are several different pricing models, such as noncooperative game and Stackelberg game shown as follows.
\subsubsection{Non-Cooperative Game}
Non-cooperative game (NCG) is a competition among individual players, in which players pursue their individual highest utilities and never consider the possibility of trying to form a coalition.
The typical radio resource market in wireless networks is given as an example to show the detailed process in the following.
In particular, there are $N$ network operators competing with each other to sell spectrum to users.
The game can be expressed as: ${\cal G}=\{\Omega, {\cal X}, \{U_n\} \}$, where $\Omega=\{1, 2, \ldots, N\}$ is the set of rational players, ${\cal X}={\cal X}_1\times\ldots\times {\cal X}_N$ is the strategy space with ${\cal X}_n$
being the strategy of player $n$, and $U_n(x_1, x_2, \ldots, x_N)$ is the utility of player $n$, which depends on the strategies of all players.
Each player $n$ is interested in maximizing its own utility $U_n(x_1, \ldots, x_N)$ via selecting spectrum price $x_n$.
Let $x_n^{*}$ be the best strategy of player $n$ which maximizes its utility.
Then, the set of the best strategies ${\boldsymbol x}^{*}=(x_1^{*}, \ldots, x_N^{*})$ is the Nash equilibrium (NE) if no player can gain higher utility by changing its own strategy when the strategies of the others remain the same.

\subsubsection{Stackelberg Game}
Stackelberg game is a sequential game in which the first mover of the strategy is the leader, and then the follower makes the corresponding strategy (best response) according to the leader's strategy.
Considering power management market in cognitive radio networks, in which the primary user (PU) is the game leaders and jointly determines their power allocation to guarantee their QoS requirement and interference price charged for the secondary user (SU) as the game follower.
In stage I, PU decides its power allocation $\pi_1$ and interference price $x_1$ to maximize its profit function $U_1(\pi_1, x_1, p_s)$, where $p_s$ denotes the power allocation demand of SU.
Subsequently, in stage II, given PU's power allocation $\pi_1$ and interference price $x_1$, the follower maximizes its utility function $U_s(p_s, \pi_1, x_1)$ by determine its power allocation $p_s$.
The objective of such a game is to find the Stackelberg equilibrium (SE).

The work in \cite{meshkati2006game} has used a non-cooperative game-theoretic approach to analyze energy-efficient power control in CDMA systems, where the objective of each user is to maximize its own utility.
The utility was computed based on the number of reliable bits transmitted over all the carriers per joule of energy consumed, which is particularly suitable for future networks.
The work in \cite{kang2012price} has investigated the pricing problem in two-tier femtocell networks, where distributed femtocells coexist with a central macrocell and share the same frequency band.
In this model the macrocell charges the femtocells with interference prices, including uniform and non-uniform pricing.
First,  assuming that the femtocells are sparsely deployed within the macrocell, the closed-form price and power allocation solutions for the formulated Stackelberg game were derived.
Then the model was extended for densely deployed case in which the cross-femtocell interference was presented and lower and upper bounds on the achievable revenue for the macrocell were obtained, as a function of price.

Naturally, by bring pricing approaches from the existing works to the resource management strategies for RIS-aided wireless networks, the diverse objectives can be reached.
Moreover, the pricing approaches only have the characteristic of partial information exchange, making them suitable for autonomous decision-making in a distributed method.
To put it crudely, in RIS-aided wireless networks, the demand for diverse objectives of a large number of entities will be reached  by exploiting pricing models.
A flexible resource management method can be developed by pricing approaches to deal with the complex and fast variations of the wireless communication environment.
Also, through negotiation mechanisms, pricing approaches enable the different entities to achieve multiple objectives such as energy consumption minimization, coverage improvement, high reliability, and fairness.

In particular for this survey, the price is commonly determined by using Stackelberg game.
This in essence attributes to the fundamental supplementary function of RIS in future wireless communications where RIS can well overcome the shortcomings of poor signal penetration in mmWave and even higher frequency communications.
For example, in the indoor environment, the transmitters and RISs can be deployed according to the requirements, so as to avoid the emergence of blocked links.
Likewise, in outdoors, cities with high-rise buildings are very easy to block the communication link.
Installing RISs on walls of buildings can improve the coverage of high-frequency communications.
6G is an era when everything is interconnected.
In this context, the RIS can carry out flexible network deployment, support flexible network architecture and various data applications, and improve the terminal experience.
On the other hand, the RIS holders and WSPs usually have their own individual benefits when the RIS is used to modify the transmission environment, and their benefits may conflict with each other, which calls for flexible and decentralized resource management strategies.
In addition to system performance and QoS requirements, from a business perspective, incentives economic factors such as cost, revenue, and profit are essential drivers to sustain RIS development and operation.
Thus, it is necessary to design incentive mechanisms whereby the WSPs pay for the RISs' services.
In the spirit of existing works in 5G \cite{luong2018applications},  Stackelberg game theory plays an important role.

Therefore, Stackelberg game methods are regarded as an alternative when designing and implementing RIS services.
The approaches aim to analyze how RIS pricing works and how RIS network entities interact economically.
In the following, we discuss important pricing approaches and RIS-aided wireless networks related works.
Among the early contributions in this areas, Ref. \cite{gao2020stackelberg} has studied the wireless resource management in RIS-aided networks given the assumption that the BS and RIS  belong to different operators.
In the spirit of this work, the authors of \cite{gao2021reflection} have investigated the {\em true} reflection resource management in RIS-aided peer-to-peer networks by defining the modular structure of RIS builds on the premise that the access points and RIS belong to different operators.

\section{Pricing Models for RIS-aided Wireless Networks}
\subsection{Stackelberg Game Formulation}

As shown in Fig. \ref{fig:3}(b2), the RIS-aided wireless network contains $S$ RIS holders that competitively forward signals generated from a BS with $M$ antennas to $K$ single-antenna mobile users.
Denote the total number of elements of RIS $s$ as $L_s$, and the reflecting coefficient is restricted by the peak power.
As discussed earlier, RIS is ubiquitous in the wireless networks,
the BS can buy use rights of the single and multiple RISs for their own applications.
Specifically, thus the BS is charged the price $q_s>0$ to lease the use rights of RIS $s.$
\begin{figure*}[!t]
\centering
\includegraphics[width=1\linewidth]{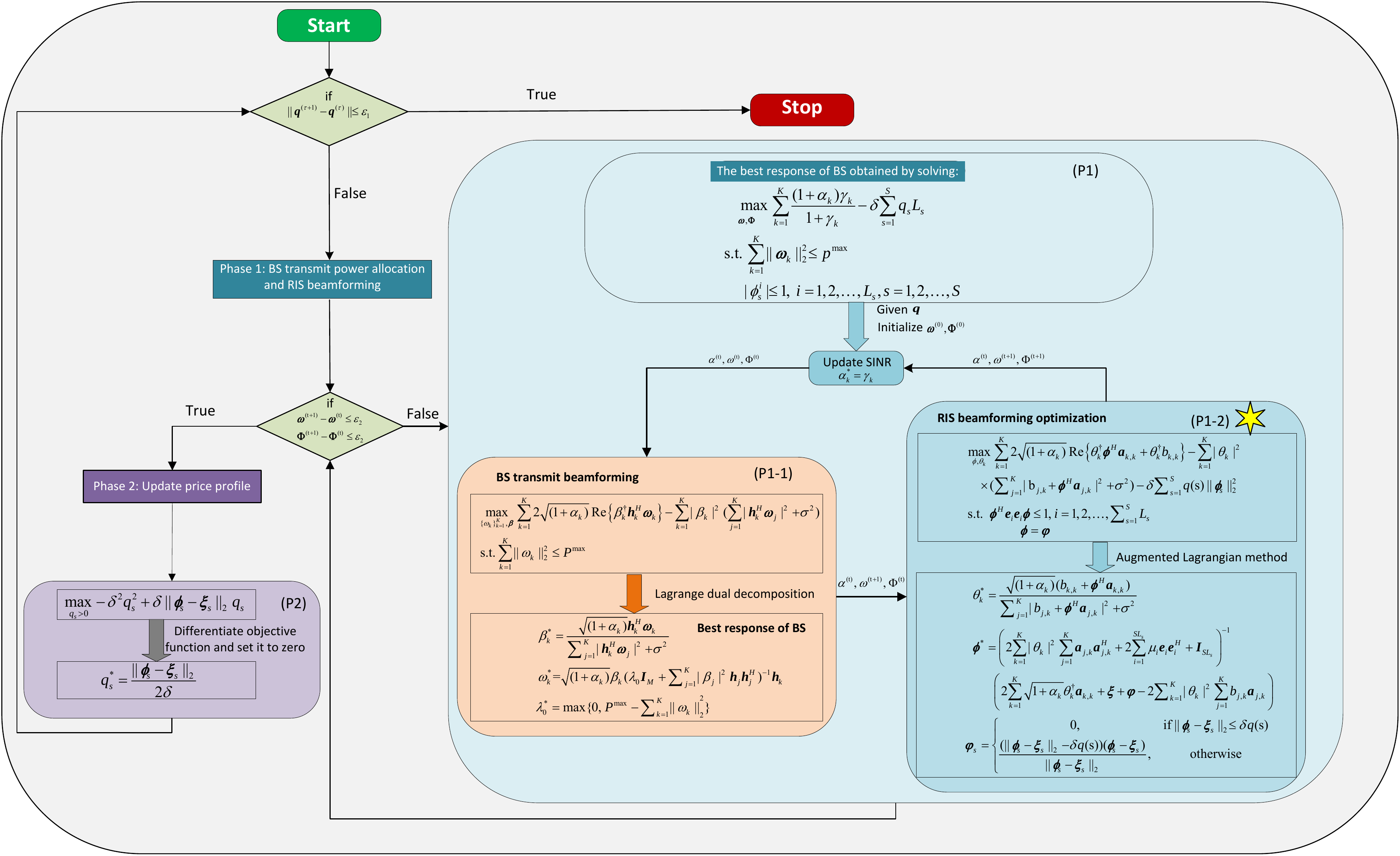}
\caption{Block diagram of backward induction algorithm.}
\label{fig:4}
\end{figure*}

Given the above RIS services, we tend to study the competition in setting the price of RIS.
For the substitute case, in this section, we present the Stackelberg game formulation for the price-based resource management.
The Stackelberg game model \cite{myerson2013game} is thus applied in this scenario.
The players, strategies, and process of Stackelberg game are presented in Fig. \ref{fig:3}.
Specifically, Stackelberg game is a strategic game that consists of leaders and followers competing with each other on certain resources.
The leaders move first and the followers move subsequently.
In this paper, we formulate the RISs as leaders, and  BSs as the followers. The RISs (leaders) impose a set of prices on per use of RIS.
Then, the BSs (followers) update their active-passive beamforming strategies to maximize their utilities perspectively based on the assigned use prices.

Under the above game model, it is easy to observe that the RIS's objective is to maximize its utility obtained from selling the use right to the BS.
Mathematically, by pricing each element, the utility of RIS $s$, that is profit, can be expressed as
\begin{equation}\label{eq:2}
V_s(q_s, {\boldsymbol q}_{-s})=q_sL_s,
\end{equation}
where ${\boldsymbol q}_{-s}$ is a vector of use prices for all RISs except RIS $s,$ i.e.,  ${\boldsymbol q}_{-s}=[q_1, \ldots, q_{s-1}, q_{s+1},\ldots, q_{S}]^T$,  $L_s$ represents the size of the $s\text{-th}$ RIS.

We let ${\pmb\Phi}\in{\mathbb C}^{L\times L}$ denote the diagonal matrix with diagonal block entries $[\pmb\Phi]_s^{\psi(s)}$ representing the phase shifts applied by RIS $s$ reflecting elements, where $\psi(s): {\cal S}\mapsto \{0, 1\}$ is a generic function that maps the sale outcome of each RIS $s$: if $\psi(s)=0$, RIS $s$ will be idle, and be bought for serving the BS otherwise.
In the following, ${\cal S}=\{1,\ldots s\ldots, S\}$ and ${\cal K}=\{1, \ldots k \ldots, K\}$ denote the sets of RIS and user, respectively, and $L=\sum_{s\in{\cal S}}L_s$ accounts the number of RIS's elements.
Let ${\boldsymbol h}_{d,k}\in{\mathbb C}^{M\times 1}$ and ${\boldsymbol G}_k:={\boldsymbol H}^H{\pmb\Phi}{\boldsymbol g}_k\in{\mathbb C}^{M\times 1}$ denote the direct and cascaded channel vectors from the BS to user $k$ with ${\boldsymbol g}_k\in{\mathbb C}^{L\times 1}$ representing the channel vector between the used RISs to user $k$, where ${\boldsymbol H}\in{\mathbb C}^{L\times M}$ is the channel matrix from the BS to the used RISs.
Moreover, the interference experienced by user $k$ from remaining users in set ${\cal K}\setminus\{k\}$ is $I_k=\sum_{i\neq k}^K|({\boldsymbol h}_{d,k}^{H}+G_k){\boldsymbol H}){\boldsymbol w}_i|^2$.
Accordingly, the SINR achieved by user $k$ takes the form:
\begin{align}
\gamma_k=\frac{|({\boldsymbol h}_{d,k}^{H}+G_k){\boldsymbol w}_k|^2}
{I_k+\sigma^2},
\end{align}
where ${\boldsymbol w}_k\in {\cal C}^{M\times 1}$ is the transmit beamforming vector for user $k$ and $\sigma^2$ is the background noise.

Hereto, the utility of the BS is defined as follows:
\begin{align}\label{eq:1}
U({\pmb\omega}_k, {\pmb\Phi}, {\boldsymbol q})=\sum\nolimits_{k=1}^K \log\left( 1+\gamma_k\right)-\delta\sum\nolimits_{s=1}^Sq_sL_s,
\end{align}
where $\delta>0$ is the weight of the cost for buying RIS $s$, and $\boldsymbol q$ is the use price vector with $\boldsymbol q=[q_1, q_2, \ldots, q_S]^T.$
It is observed from (3) that the utility function of the BS consists of two parts: {\em profit} and {\em cost}.
If the BS increases its transmit power and the size of the RISs, the transmission rate increases, and so does the profit.
As a result, it has to buy more RISs services, which increases the cost.
Therefore, resource management strategies are needed at the BS to maximize their own utilities.

The objective of this Stackelberg game is to find the SE solution from which neither the leaders nor the followers have incentives to deviate.
For any $({\boldsymbol q}, {\pmb\omega}, {\pmb\Phi})$, the SE solution $({\boldsymbol q}^{*}, {\pmb\omega}^{*}, {\pmb\Phi}^{*})$  should comply with
$U({\boldsymbol q}^{*}, {\pmb\omega}^{*}, {\pmb\Phi}^{*})\geq U({\boldsymbol q}^{*}, {\pmb\omega}, {\pmb\Phi})$ and
$V_s({\boldsymbol q}^{*}, {\pmb\omega}^{*}, {\pmb\Phi}^{*})\geq V_s(q(s)^{*}, {\boldsymbol q}_{-s}^{*}, {\pmb\omega}^{*}, {\pmb\Phi}^{*}).$
As mentioned in \cite{aumann1995backward}, the backward induction method is commonly used to compute the SE.

We apply this Stackelberg game structure to obtain the equilibrium of active-passive beamforming and prices for RIS holders.
With an assumption that the BS and RIS holders are rational to maximize their profits, we present the backward induction algorithm  for solving the Stackelberg game problem.
The algorithm consists of two phases: (i) {\em BS transmit power allocation and RIS beamforming} phase, (ii) {\em prices update } phase as illustrated in Fig. \ref{fig:4}.
In phase 1, we firstly apply the Lagrangian dual transform to tackle the logarithm in the BS's utility function.
Then, given price ${\boldsymbol q}(\tau)$, the BS transmit power allocation and RIS beamforming can be obtained by solving the equivalent optimizing problem (P1) as presented in Fig. \ref{fig:4}.
In order to develop a tractable algorithm for (P1), a convenient approach is to employ the alternating optimization technique to separatively and iteratively solve for $\pmb\omega$ and $\pmb\Phi$.
More precisely, in each iteration, we first update the nominal SINR $\pmb\alpha$, and then better solution for $\pmb\omega$ and $\pmb\Phi$ is updated by addressing (P1-1) and (P1-2), respectively.
Based on the best response of the BS, the RIS holders can adjust the price ${\boldsymbol q}$ to achieve the highest profit.
In addition, all the control variables and auxiliary variables are initialized. We introduce the auxiliary variables as ${\pmb \alpha}=[\alpha_1, \alpha_2, \ldots, \alpha_K]^T$, $\pmb\beta=[\beta_1, \ldots, \beta_K]$, $\pmb\theta=[\theta_1, \ldots, \theta_K]$, $\lambda_0$, $\pmb\varphi=[\pmb\varphi_1, \ldots, \pmb\varphi_s, \ldots, \pmb\varphi_S]$, $\varepsilon_1$, and $\varepsilon_2$.
Here,  $\alpha_k$ is used to decode SINR $\gamma_k$, $\varepsilon_1$ and $\varepsilon_2$ are used to control the stopping time of algorithm.
Auxiliary variables $\beta_k$ and $\pmb\varphi_s$ are introduced to deal with the multiple-ratio fractional programming problem (P1-1) and (P1-2), respectively.
Moreover, $\lambda_0$ and $\pmb\varphi$ are the corresponding Lagrangian multipliers for BS power budget of (P1-1) and reflection coefficient constraints of (P1-2), respectively.

As shown in Fig. \ref{fig:4}, in each iteration step for solving (P1), the highest complexity is for finding ${\pmb \phi}^{*}$, in which the complexity of the summation operation, the matrix inversion, and the final matrix multiplication is quantified by ${\cal O}(\sum_{s\in{\cal S}}\psi(s)L_s)$, ${\cal O}([\sum_{s\in{\cal S}}\psi(s)L_s]^3)$, and ${\cal O}([\sum_{s\in{\cal S}}\psi(s)L_s]^2)$, respectively.
Thus, the complexity of the Lagrangian dual decomposition is ${\cal O}([\sum_{s\in{\cal S}}\psi(s)L_s]^6).$
Obviously, the density of RISs deployment and a large number of reflection elements installed in each RIS seriously impact the scalability of the proposed algorithm.
Therefore, it is worthy designing low-complexity algorithm to this Lagrangian dual decomposition method.
Continuing the previous work \cite{gao2021reflection}, the modular  RIS can be fully utilized in the proposed pricing model for RIS-aided wireless networks, which is expected to be a new research direction in future.

\subsection{Numerical Results}
In this subsection, the performance of pricing-based resource management in an RIS-aided wireless network is evaluated, relying on the Stackelberg game-based method developed in the last section. We consider an RIS-aided wireless network for both the BS and $5$ RISs as well as $4$ single-antenna users.
The BS with $4 $ antennas is deployed at $(0,0)\text{~m}$, and users are uniformly distributed within a circle, whose size and locations are prescribed by its radius $10\text{~m}$  and coordinate $(200, 0)\text{~m}$.
Furthermore, for the convenience to describe the distribution of $5$ RISs, we construct a diamond, in which two diagonals are parallel to the horizontal and vertical respectively.
More precisely, the horizontal and vertical diagonal lengths are $25\text{~m}$ and $50\text{~m}$, respectively.
The locations of $5$ RISs are described as follows: RIS $5$ is deployed at the intersection of diagonal lines and the remaining RISs fall from the top of the diamond counterclockwise in an ascending order.
In particular, we assume RIS 3 located at $(50, 0)\text{~m}$ except for special instruction.
As for the communications channel, we consider both the small scale fading and the large scale path loss.
Specifically, the path loss exponent of the link between the BS and the user, that of the link between the user and RISs, as well as that of the link between RISs and the BS are $3.5$,  and $2$, respectively, and the path loss at the reference distance $1\text{~m}$ is set as $30\text{~dBm}$ for each individual link, while the small scale fading is accounted by Rayleigh fading model \cite{gao2021reflection}.
\begin{figure*}[!t]
\centering
\includegraphics[width=1\linewidth]{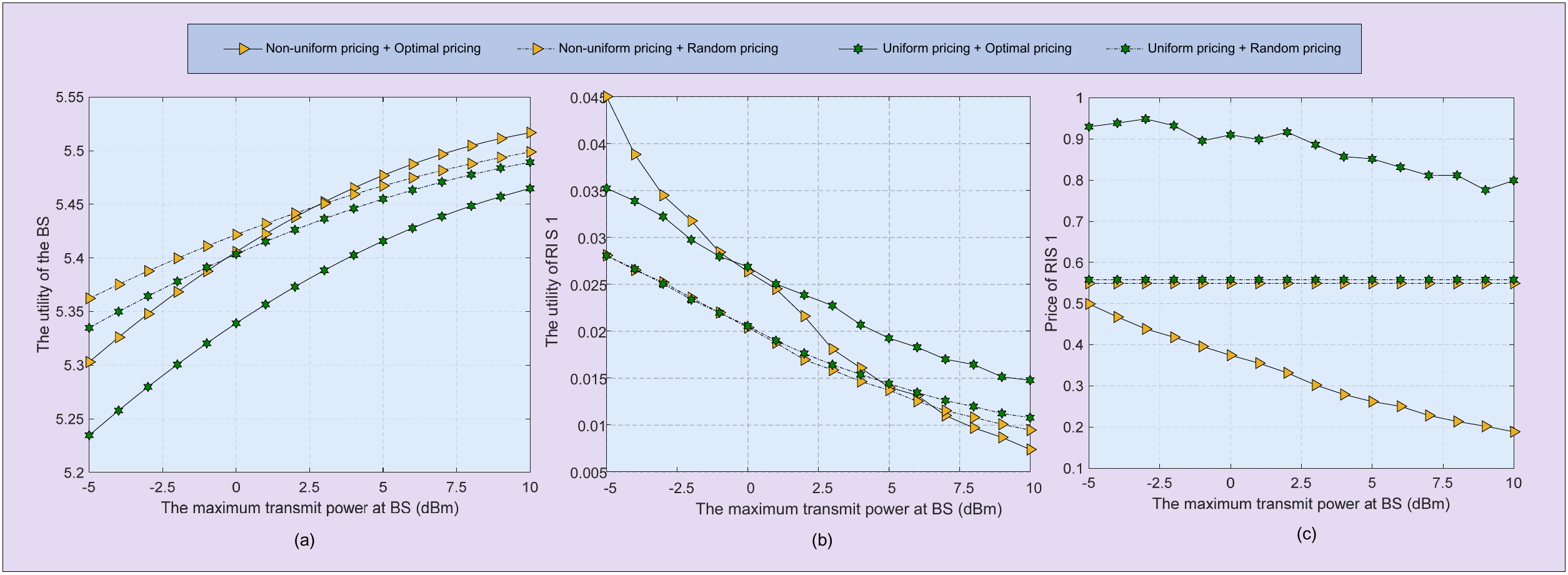}
\caption{a) Utility of the BS, b) utility of RIS 1, and c) price of RIS 1 as  functions of BS power budget.}
\label{fig:5}
\end{figure*}

The following  details our simulation results, in terms of the utility performance in both the WSPs and the RIS holders in the proposed Stackelberg game-based and random pricing methods for various simulation environments.
The following two pricing schemes are considered: {\em uniform pricing} and {\em non-uniform pricing}.
To quantify the impact of power budget,  Fig. \ref{fig:5} depicts the utilities of the BS as well as the utility performance and price of RIS 1  versus power budget of the BS, for various pricing schemes.
Our observations are as follows.
For all these four schemes after combination, the increase of transmit power budget drastically augments the utility of the BS.
This is because the SINR part of the BS's utility imposed by the transmit power dominates when the power budget is of a large value.
Meanwhile, it can be readily observed from Fig. \ref{fig:5}(b) that the utility value of  RIS 1 achieved by the Stackelberg game-based scheme decreases with the increase of power budget.
This is partially caused by the reduced size of RISs used (purchased) by the BS and partially due to the reduced pricing of each RIS.
In fact, this in essence attributes to the construction of the utility function of the BS, in which the cost of power consumption is not considered, and thereby, the BS will tend to buy a small number of RISs when the transmit power is sufficient.
Consequently, for $p^{\max}>0\text{~dBm},$  we observe that RIS 1 attempts to incentive the BS to purchase reflection resource at low price, which can be observed from Fig. \ref{fig:5}(c).

\begin{figure}[!t]
\centering
\includegraphics[width=1\linewidth]{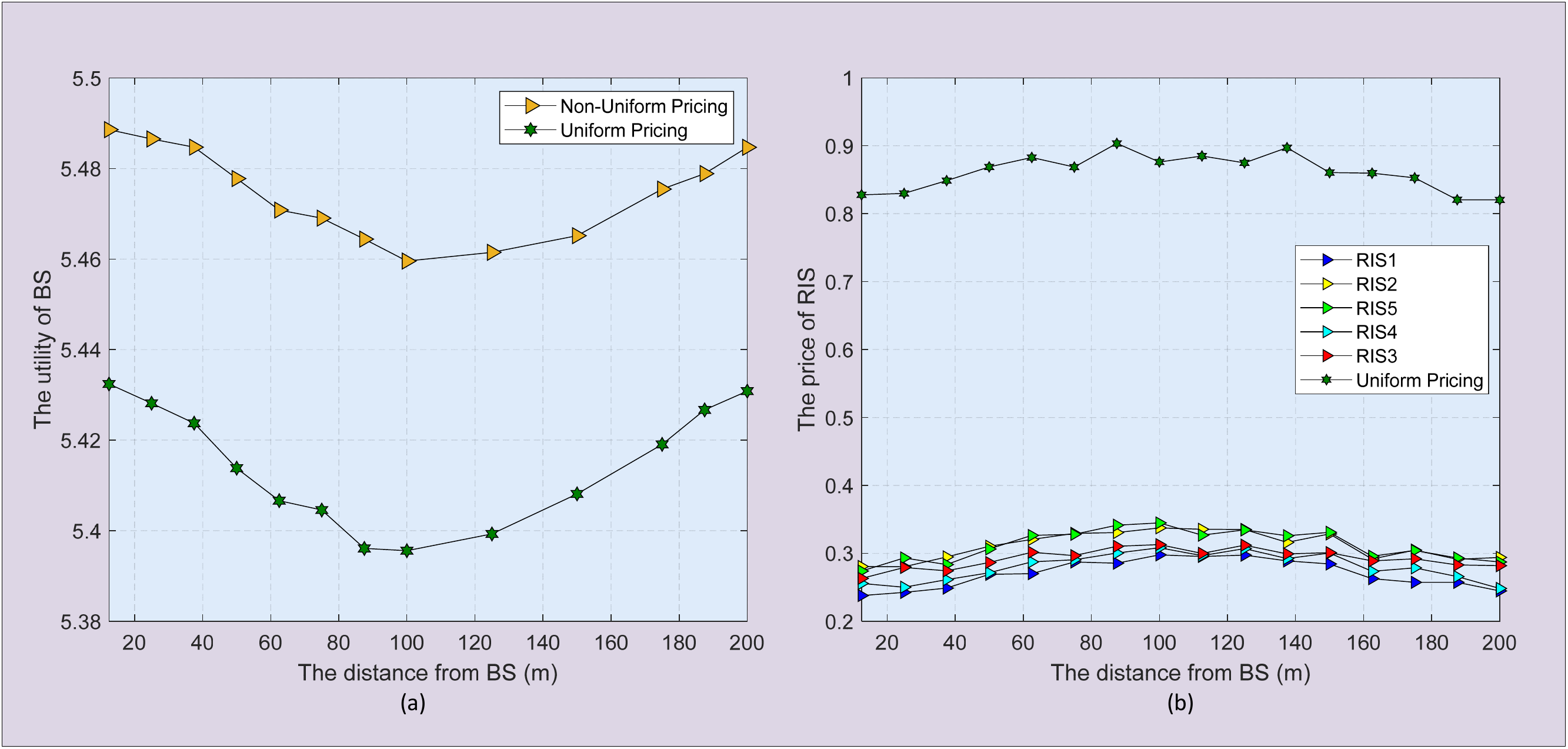}
\caption{a) Utility of the BS and b) price at the equilibrium versus the location of RISs. }
\label{fig:6}
\end{figure}

Then, we investigate the impact of RISs' deployment locations.
For simplification, the diamond deployed with $5$ RISs swims along the $x\text{-direction}$ as a whole, we move the intersection of this diamond from $(12.5, 0)\text{~m}$ to $(200, 0)\text{~m}$, and respectively plot the utility of the BS and RISs' prices with respect to the distance from the BS in Fig. \ref{fig:6}(a) and \ref{fig:6}(b),  while setting the power budget of the BS as $10\text{~dBm}.$
As can be observed from Fig. \ref{fig:6}(a),  the BS's utility performance of uniform-pricing and non-uniform pricing first decreases and then increases with the distance between RIS 5 and the BS.
This in  essence attribute to the double-fading path-loss model in SINR $\gamma_k$.
In contrast, from Fig. \ref{fig:6}(b), we observe that the prices of RISs first slightly increase and then decrease while increasing the distance between RIS 5 and the BS.
In fact,  for any given value of $K, S, L_s$, and power budget, it can be also inferred from the double-fading path-loss model that using more RIS elements can increase the utility of the BS, especially when the cluster of RISs is deployed far away from the BS or the users' cluster.
Correspondingly, for the cluster of RISs, the RIS's price is negatively correlated to the cascade channel gain of BS-RIS-User.
\subsection{Future Research Directions}

As observed, pricing for RIS-aided wireless networks is a meaningful research area.
Since it is an emerging topic, there are still many open issues to be addressed, some of which are listed as follows.
\begin{itemize}
\item {\bf Community Pricing:} It is necessary to investigate the {\em community pricing} for the RIS-aided wireless networks in complex urban environment, since there are a large number of residents in each community.
    To be more specific,  during the construction of communities, the community developers can price the deployment of RISs rather than householders.
    Community pricing can well reduce the complexity, since all the RISs in one community belong to one holder.
\item {\bf Function Pricing:} Function pricing is needed to meet diverse user requirements in different communication scenarios.
    The function pricing is based on serving different communication needs, which can provide guidance of the deployment and size of RIS.
\item {\bf Vickrey Auction Pricing:} With the increasing contradiction between the shortage of wireless resources and soaring performance requirements, the on-demand resource allocation is significantly important.
    Fortunately, auction is an efficient way of scheduling resources to buyers which value the resources  most.
    In the Vickrey action, WSPs (bidders) will bid prices since they are willing to pay for using RISs to the auctioneer.
    The highest bidder will win, as determined by the auctioneer.
    In the end, the winner will pay the second-highest price rather than his own submission.
\end{itemize}

\section{Conclusion}

RIS has emerged as one of the  promising technologies for 6G wireless networks.
In this paper we have considered the pricing approaches for resource management in RIS-aided wireless networks.
Firstly, we have described the principles and some typical RIS applications in various emerging systems.
Then, we have specifically introduced the heterogeneous characteristics of RIS networks with the aim to understand the motivations of using pricing models in RIS-aided networks.
Afterwards, to demonstrate the application of pricing model, we presented the Stackelberg game theoretic model for RISs service competition.


\end{document}